\documentstyle[prl,aps,epsf]{revtex}
\draft
\begin{document}
\twocolumn[\hsize\textwidth\columnwidth\hsize\csname @twocolumnfalse\endcsname

\title{Metastability and Transient Effects in Vortex Matter
Near a Decoupling Transition}
\author{C.~J. Olson and C. Reichhardt}
\address{Theoretical Division and Center for Nonlinear Studies,
Los Alamos National Laboratory, Los Alamos, New Mexico 87545.}

\author{R.~T.~Scalettar and G.~T.~Zim{\' a}nyi}
\address{Department of Physics, University of California, Davis, California
95616.}

\author{Niels Gr{\o}nbech-Jensen}
\address{Department of Applied Science, University of California,
Davis, California 95616.\\
NERSC, Lawrence Berkeley National Laboratory, Berkeley, California 94720.}

\date{\today}
\maketitle
\begin{abstract} 
We examine metastable  
and transient effects both above and below the first-order 
decoupling line in a 3D simulation of magnetically interacting
pancake vortices. We observe pronounced transient and history effects  
as well as 
supercooling and superheating between the 
3D coupled, ordered and 2D decoupled, disordered phases.
In the disordered supercooled state 
as a function of DC driving, reordering occurs through the
formation of growing moving channels of the ordered phase. 
No channels form in the superheated region; instead the ordered
state is homogeneously destroyed.
When a sequence of current pulses is applied we observe memory effects.
We find a ramp rate dependence of the $V(I)$ curves on both sides
of the decoupling transition.  The critical current that we obtain depends
on how the system is prepared.
\end{abstract} 

\vskip2pc]
\narrowtext

Vortices in superconductors represent an ideal system in which to study
the effect of quenched disorder on elastic media.
The competition between 
the flux-line interactions, which order the vortex lattice, and the defects in
the sample, which disorder the vortex lattice, produce a remarkable 
variety of collective behavior \cite{Review1}. 
One prominent example is the peak effect in low temperature 
superconductors, which appears near $H_{c2}$ when a
transition from an ordered to a strongly pinned disordered state
occurs in the vortex lattice 
\cite{Kes2,Shobo3,Chaddah4,Henderson5,Andrei6,Xiao7,Paltiel8}.  
In high temperature superconductors, particularly BSCCO samples, a striking 
``second peak'' phenomenon is observed  
in which a dramatic increase in the critical current occurs 
for increasing fields.  It has been proposed that this is an
order-disorder or 3D to 2D transition.  
\cite{Giamarchi9,Cubitt10,Tamegai11,Glazman12,Feigelman13}

Recently there has been renewed interest in
transient effects, which have been observed in voltage response
versus time curves in low temperature superconductors 
\cite{Henderson5,Andrei6,Xiao7,Frindt14,Baker14a,Good14b,Anjaneyulu14c}. 
In these experiments the voltage response 
increases or decays with time, depending on how the vortex lattice
was prepared. The existence of
transient states suggests that the disordered phase can be {\it supercooled} 
into the ordered region
\cite{Chaddah14d}, producing an increasing voltage response, whereas
the ordered phase may be {\it superheated} into the disordered region, giving
a decaying response. 
In addition to transient effects,
pronounced memory effects and hysteretic V(I) curves have been
observed near the peak effect in low temperature materials
\cite{Kes2,Chaddah4,Henderson5,Andrei6,Xiao7,Good14b,Anjaneyulu14c,Henderson14c,Banerjee14c,Ravikumar14c}.
Memory effects are also seen in simulations \cite{jensen}.
Xiao {\it et al.} \cite{Xiao7} 
have shown that transient behavior can lead
to a strong dependence of the critical current on the current
ramp rate. 
Recent neutron scattering experiments 
in conjunction with ac shaking have provided more direct 
evidence of supercooling and superheating near the peak effect 
\cite{Ling15}. Experiments on BSCCO have revealed that the high field 
disordered state can be supercooled to fields well below the
second peak line \cite{Konczykowski16}. 
Furthermore, transport experiments in BSCCO have shown
metastability in the zero-field-cooled state near the second peak 
as well as hysteretic V(I) curves \cite{Portier17}, and
magnetooptic imaging has revealed the coexistence of ordered and 
disordered phases \cite{Giller18}.
Hysteretic and memory effects have also been
observed near the second peak in YBCO 
\cite{Kokkaliaris19,Bekeris20,Esquinazi21}.   

The presence of metastable states and superheating/supercooling effects
strongly suggests
that the order-disorder transitions in these different 
materials
are {\it first order} in nature. 
The many similarities
also point to a universal behavior between the peak effect
of low temperature superconductors and the peak effect and
second peak effect of high temperature superconductors. 

A key question in all these systems is the nature of the
{\it microscopic dynamics} of the vortices 
in the transient states; particularly, whether plasticity or 
the opening of flowing channels are involved \cite{Andrei6}. 
The recent experiments have made it clear that a proper 
characterization of the static and dynamic phase diagrams must
take into account these metastable states, and therefore an understanding
of these effects at a microscopic level is crucial.  
Despite the growing amount of experimental work on  
metastability and transient effects in vortex matter near
the peak effect transition \cite{Henderson5,Andrei6,Xiao7}, 
these effects have not yet been 
investigated numerically.  

In this work 
we numerically study magnetically interacting
pancake vortices driven through quenched point disorder. 
As a function of 
applied field, temperature, or interlayer coupling
the model exhibits a sharp 3D (coupled, ordered phase) to 
2D (decoupled, disordered phase) 
transition, consistent with
theoretical expectations \cite{Glazman12,Feigelman13},
that is associated with a large change in the
critical current \cite{FirstPaper22}. 
Near the disordering transition, we find strong metastability and transient
effects.
A metastable state is a thermodynamic state that is out of
equilibrium, which persists for a time longer than the characteristic
relaxation time of the system at equilibrium \cite{Binder22a}.
By supercooling or superheating the ordered and disordered phases,
we find increasing or decreasing
transient voltage response curves,
depending on the amplitude of the drive pulse and the proximity to the 
disordering transition. 
In the supercooled transient states a
growing ordered channel of flowing vortices forms.
No channels form in the
superheated region but instead the ordered state is homogeneously destroyed.
We observe memory effects when a sequence of pulses is applied,
as well as ramp rate dependence and hysteresis in the V(I) curves. 
The critical current we obtain depends on how the system is prepared.

We consider a 3D layered superconductor containing an equal number of
pancake vortices in each layer, interacting magnetically.
We neglect the Josephson coupling, which is
a reasonable approximation for highly anisotropic materials.
The overdamped equation of motion for vortex $i$ at $T=0$ is 
\begin{equation}
{\bf f}_{i} = -\sum_{j=1}^{N_{v}}\nabla_{i} {\bf U}(\rho_{ij},z_{ij})
+ {\bf f}_{i}^{vp} + {\bf f}_{d} = \eta {\bf v}_{i}.
\end{equation}
The total number of 
pancakes is $N_{v}$, and $\rho_{ij}$ and $z_{ij}$ are the distance between
vortex $i$ and vortex $j$ in cylindrical coordinates. We impose 
periodic boundary conditions in the $x$ and $y$ directions and open
boundaries in the $z$ direction. The magnetic interaction energy
between pancakes is \cite{Clem22a,Brandt23}
\begin{eqnarray}
{\bf U}(\rho_{ij},0)=2d\epsilon_{0}
\left((1-\frac{d}{2\lambda})\ln{\frac{R}{\rho}}
+\frac{d}{2\lambda}
E_{1}\right) \ ,
\end{eqnarray}
\begin{eqnarray}
{\bf U}(\rho_{ij},z)=-s_{m}\frac{d^{2}\epsilon_{0}}{\lambda}
\left(\exp(-z/\lambda)\ln\frac{R}{\rho}+
E_{2}\right) \ , 
\end{eqnarray}
where 
$E_{1} = 
\int^{\infty}_{\rho} d\rho^{\prime} 
\exp(-\rho^{\prime}/\lambda)/\rho^{\prime}$,
$E_{2} = 
\int^{\infty}_{\rho} d\rho^{\prime} 
\exp(-\sqrt{z^{2}+\rho^{\prime 2}}/\lambda)/\rho^{\prime}$,
$R = 22.6\lambda$, the maximum in-plane distance, 
$\epsilon_{0} = \Phi_{0}^{2}/(4\pi\lambda)^{2}$, $d=0.005\lambda$ is the
interlayer spacing as in BSCCO, 
$s_{m}$ is the coupling strength, and
$\lambda$ is the London penetration depth. 
The viscosity $\eta=B_{c2} \Phi_{0}/\rho_{N}$, where
$\rho_{N}$ is the normal-state resistivity.  Time is
measured in units of $\eta/f_{0}^{*}$.
For 

\begin{figure}
\center{
\epsfxsize = 3.5 in
\epsfbox{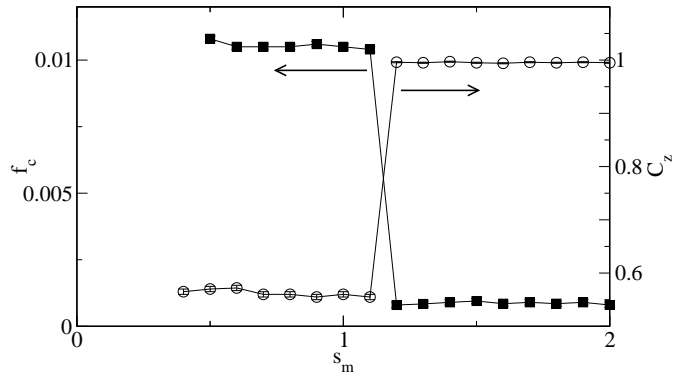}}
\caption{
Critical current $f_c$ (filled squares) and 
interlayer correlation $C_z$ (open circles)
for varying $s_m$ in a system with 16 layers, $n_{v}=0.3/\lambda^{2}$, 
and $f_{p}=0.02f_{0}^{*}$,
showing the sharp transition from coupled behavior at 
$s_{m} \le 1.2$ to decoupled behavior at $s_{m} > 1.2$.
}
\label{fig:fc}
\end{figure}

\noindent
example, 
in the case of a BSCCO sample 
2 $\mu$m thick, taking
$\lambda = 250$ nm, $\xi = 1.8$ nm, and $\rho_N = 2.8$ m$\Omega$ cm
gives the value of a single time step as 0.4 ns.
We model the pinning
as $N_p$ short range attractive 
parabolic traps that are randomly distributed in each 
layer. 
The pinning interaction is 
${\bf f}_{i}^{vp} = -\sum_{k=1}^{N_{p}}(f_{p}/\xi_{p})
({\bf r}_{i} - {\bf r}_{k}^{(p)})\Theta( 
(\xi_{p} - |{\bf r}_{i} - {\bf r}_{k}^{(p)} |)/\lambda)$,
where the pin radius 
$\xi_{p}=0.2\lambda$, the pinning force  
is $f_{p}=0.02f_{0}^{*}$, and $f_{0}^{*}=\epsilon_{0}/\lambda$.
We fix the temperature $T=0$.
To vary the applied magnetic field $H$, we fix the number of vortices
in the system and change the system size, thereby 
changing the vortex
density $n_v$.  The pin density remains fixed.
We use $L=16$ layers throughout 
this work, and 
except where mentioned, we focus on a
system that ranges from $12.9\lambda \times 12.9\lambda$
to $14.8\lambda \times 14.8\lambda$, with
a pin density of $n_p = 1.0/\lambda^{2}$ 
in each of the layers. 
We have also studied a sample with stronger, denser pinning of 
$\xi_{p}=0.1\lambda$,
$f_{p}=0.08f_{0}^{*}$, and $n_{p}=8.0/\lambda^{2}$,
of size $5.1\lambda \times 5.1\lambda$ to $6.9\lambda \times 6.9 \lambda$.
In each case there are $N_{v}=80$ vortices per layer, giving 
a total of 1280 pancake vortices.

For sufficiently strong disorder, the vortices 
in this model show a sharp 3D-2D
decoupling transition as a 
function of coupling strength $s_m$,
vortex density $H$ \cite{FirstPaper22,PhysicaC22b},
or temperature \cite{PhysicaC22b,Vinokur22a}.
A dynamic 2D-3D transition can also occur 
\cite{FirstPaper22}.
There are actually two disordering transitions that occur in the
model: a decoupling transition from 3D to 2D, and an in-plane disordering
transition.  In our studies, we find that these two transitions
always coincide, and appear as a single transition from a 3D state that
is ordered in the plane and coupled between planes, to a 2D state that is
decoupled and disordered in the plane.
We denote the magnetic field at which the static 3D-2D transition occurs
as $n_{v}^{c}$, and the coupling strength at which the transition occurs
as $s_{m}^{c}$. 
For the main system considered here,
a transition from ordered 3D flux lines 
to disordered, decoupled 2D pancakes occurs at 
$s_{m}^{c}=1.2$ with $n_{v}=0.3/\lambda^{2}$ or at 
$n_{v}^{c}=0.38$ with $s_{m}=0.7$. 
The coupling/decoupling transition occurs twice as a function
of field \cite{Paltiel8}, once at low 

\begin{figure}
\center{
\epsfxsize = 3.5 in
\epsfbox{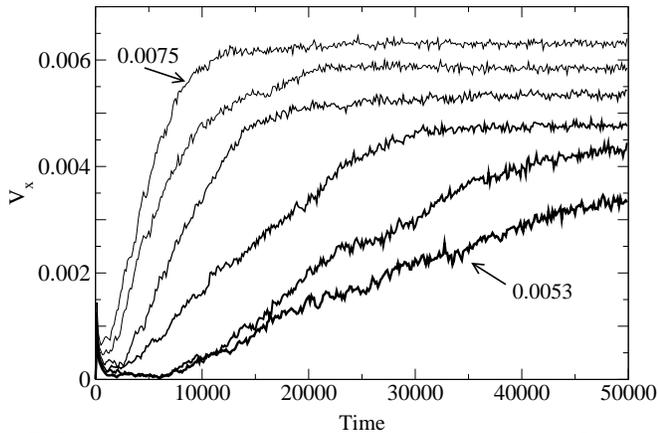}}
\caption{Increasing transient voltage response curves $V_{x}$ versus time
for supercooled state, for $s_{m}=2.0$ and $f_{d}/f_{0}^{*} =$
(bottom) 0.0053, 0.0055, 0.006, 0.0065, 0.007, and 0.0075 (top).
$V_{x}=0$ for all times for $f_{d}/f_{0}^{*}<0.0053$.
}
\label{fig:newfig2}
\end{figure}

\noindent
fields when the vortex
lines form as the effective pinning strength begins to decrease away
from the single vortex pinning regime,
and a second time at higher fields when the magnetic interactions
among vortex pancakes in a given plane cause the pancakes in different
planes to decouple.  We examine the higher field transition
in the sample with strong, dense pinning, where $n_{v}^{c}=3.0$
at $s_{m}=0.7$.
The same effects described here also 
appear on either side of
a temperature-induced 3D-2D transition.

We illustrate the difference in critical current between 
the
coupled and decoupled vortex phases in Fig.~\ref{fig:fc}.  As a function of
interlayer coupling $s_{m}$, we show the critical current
$f_{c}$, obtained by summing $V_{x}=(1/N_{v})\sum_{1}^{N_{v}}v_{x}$
and identifying the drive $f_{d}$ at which $V_{x} > 0.0005$.
Also plotted is a measure of the $z$-axis correlation,
$C_z = 1 - \left<(|{\bf r}_{i,L} - {\bf r}_{i,L+1}|/(a_0/2)) \
\Theta(a_0/2 - |{\bf r}_{i,L} - {\bf r}_{i,L+1}|\right>$, 
where $a_0$ is the vortex lattice constant, and the average is
taken over all pancakes in the system.
The ordered phase has a much lower critical current,
$f_{c}^{o} = 0.0008f_{0}^{*}$, than the 
disordered phase,
$f_{c}^{do} = 0.0105f_{0}^{*}$.

To observe transient effects, we supercool the lattice by 
annealing the system at $s_{m} < s_{m}^{c}$ into a disordered,
decoupled configuration.
Starting from this state, we set $s_m > s_{m}^{c}$
such that the pancakes would be ordered and coupled at equilibrium,
and at $t=0$ we apply a fixed drive $f_d$ for 400000 steps.
Application of a driving current is only one possible way in which
the equilibrium configuration can be regained.  The equilibrium
state can also be reached via thermal fluctuations, through gradients
in the magnetic field \cite{Konczykowski16}, or by applying a rapidly
fluctuating magnetic field \cite{Avraham22a}.
In Fig.~\ref{fig:newfig2} we show the time-dependent voltage 
response
$V_x$ for several different
drives $f_d$ for a sample which would be coupled in equilibrium, 
with $s_{m} = 2.0$, 
that has been prepared in a decoupled state at $s_{m}=0.5$. For 
$f_{d} < 0.0053 \pm 0.0001 f_{0}^{*}$ the system remains pinned 
in a decoupled disordered state.
For $f_{d} \gtrsim 0.0053 f_{0}^{*}$ 
a time dependent increasing 

\begin{figure}
\center{
\epsfxsize = 3.5in
\epsfbox{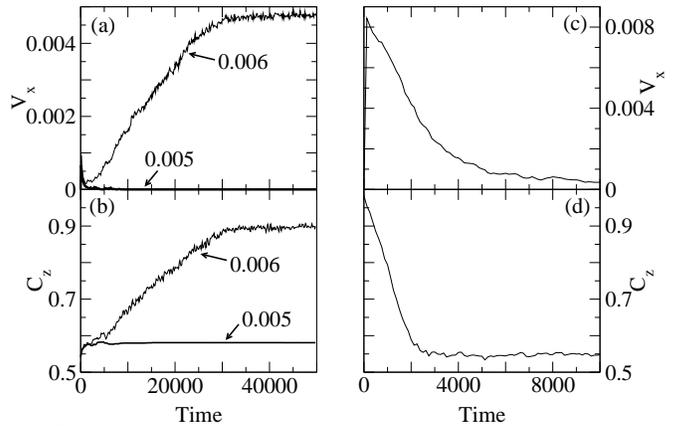}}
\caption{(a) and (c): Transient voltage response curves $V_{x}$ versus
time;  (b) and (d): corresponding transient interlayer correlation
response $C_{z}$, for a superheated and a supercooled sample.
Panels (a) and (b) show the increasing response of a 
supercooled sample with $s_{m}=2.0$
and (upper light line) $f_{d}=0.006f_{0}^{*}$,
(lower heavy line) $f_{d}=0.005f_{0}^{*}$.  Panels (c) and (d) show the 
decreasing response of a superheated
sample with $s_{m}=0.7$ and $f_{d}=0.010f_{0}^{*}$.}
\label{fig:newfig3}
\end{figure}

\noindent
response occurs. $V_x$ does not rise
instantly but only after a specific waiting time $t_{w}$.
The 
rate of increase in $V_x$ grows as the amplitude of the
$f_{d}$ increases.  As shown in Fig.~\ref{fig:newfig3}(a,b), the
$z$-axis correlation $C_{z}$
exhibits the same behavior as $V_x$, indicating that the 
vortices
are becoming more aligned in the $z$ direction as time passes.
If the vortices move in response to thermal activation,
a voltage response of the form 
$V_{x}(t)=V_{0} (1-e^{-Wt})$ should appear \cite{Baker14a}.
Since our simulation is at $T=0$, we would not expect
thermal activation to apply.
Instead, the response we observe can be fit by an exponential form
only at long times, such as $t \gtrsim 30000$ in 
Fig.~\ref{fig:newfig3}, when $V_{x}$ is beginning to saturate.
At intermediate times ($5000 < t < 30000$ in Fig.~\ref{fig:newfig3}(a,b))
the increase in $V_{x}$
and $C_{z}$ is roughly linear with time.
As we will show below, this linear increase indicates that an ordered
regime is growing at a constant rate.
The amount of time required for the system to reach a steady voltage
response level is indicative of the fact that we have started the
system in a metastable state.  If we prepare the lattice in its
equilibrium configuration of coupled lines and then apply the same
currents shown in Fig. 2, the voltage response reaches its full, steady
value within less than 100 steps, whereas in Fig. 2, 10000 to 50000 steps
are required.

\begin{figure}
\center{
\epsfxsize = 3.5 in
\epsfbox{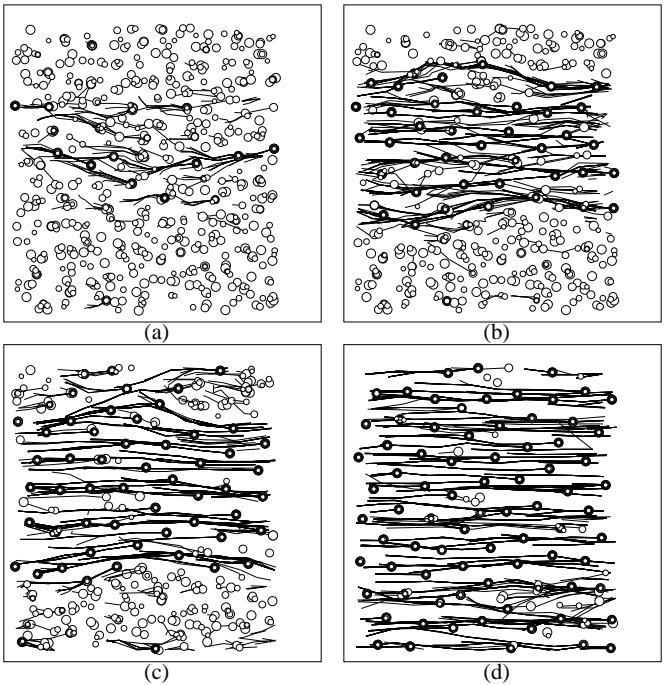}}
\caption{Pancake vortex positions (circles) and trajectories (lines) for
the supercooled phase at $s_{m}=2.0$ and $f_{d}=0.007f_{0}^{*}$, seen
from the top of the sample, at times: (a) $t=2500$; (b) $t=7500$;
(c) $t=12500$; (d) $t=20000.$  Pancakes on a given layer are represented
by circles with a fixed radius; the radii increase from the top layer to
the bottom.  An ordered channel forms in the sample and grows outward.
Coexistence of the ordered and disordered phases occurs in panels
(a) through (c).
}
\label{fig:coolI}
\end{figure}

To determine how the vortex lattice is moving 
when $V_{x}$ is nearly zero (during $t_{w}$), linearly increasing,
and saturating exponentially,
we show the vortex positions 
and trajectories in the supercooled sample 
in Fig.~\ref{fig:coolI}. 
Here a series of images have been taken from a sample 
in Fig.~\ref{fig:newfig2} with $s_m=2.0$ for $f_{d}=0.007f_{0}^{*}$ for
different times.  In 
Fig.~\ref{fig:coolI}(a) at $t=2500$ the initial
state is disordered.  In Fig.~\ref{fig:coolI}(b) at $t=7500$ significant
vortex motion occurs through the {\it nucleation} of a single channel
of moving vortices.  At lower drives the channel gradually 
appears 

\noindent
during the waiting time $t_w$, but at $f_{d}=0.007f_{0}$ 
the channel nucleates relatively
rapidly.  Vortices outside the channel remain pinned.  In 
Fig.~\ref{fig:coolI}(c) at $t=12500$ the channel is wider, and vortices
inside the channel are ordered and have recoupled.  The pinned
vortices remain in the disordered state.  During the transient motion there
is a {\it coexistence} of ordered and disordered states.  
If the drive
is shut off the ordered domain is pinned but remains 
ordered, and when
the drive is re-applied the ordered domain moves again.  
The vortices inside the ordered channel tend to align their lattice
vector in the direction of the drive \cite{drivenreorder22a}.
The increase in both $V_x$ and $C_z$ is roughly linear in time
during the period when the fully formed channel is expanding outward.
Since the velocity of the moving vortices is constant, a linear increase
in $V_x$ indicates that the {\it number} of moving vortices is also
increasing linearly in time.  Vortices do not begin to move until the
edge of the ordered channel reaches them.  Therefore, the linear increase
in $V_x$ indicates that the outward growth of the ordered channel is 
proceeding at a constant rate.
In 
Fig.~\ref{fig:coolI}(d) for $t=20000$ almost all of the vortices have
reordered and the channel width is the size of the sample.  After this
time slow rearrangements of the vortices into a more ordered configuration
occur, and 

\begin{figure}
\center{
\epsfxsize=3.5in
\epsfbox{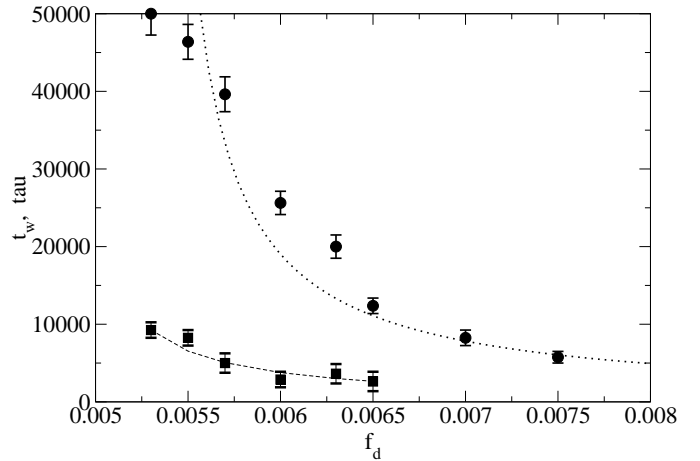}}
\caption{Squares (lower curve): 
Waiting time $t_{w}$ as a function of applied drive $f_{d}$ for
the superheated sample illustrated in Fig.~\ref{fig:newfig2}.
Dashed line indicates a fit to $t_{w} \propto 1/f_{d}$.
Circles (upper curve): 
Response saturation time $t_{s}$ for the same sample.
Dotted line indicates a fit to $t_{s} \propto (f_{d}-0.0053)^{-1}$.
}
\label{fig:twait}
\end{figure}

\noindent
$V_{x}$ switches over to an exponential saturation
with time.
Thus in
the supercooled case we observe {\it nucleation} of a microscopic
transport channel, followed by {\it expansion} of the channel.
We note that
in recent scanning Hall probe \cite{Marchevsky23d} and 
magneto-optic \cite{Giller23e} experiments, coexisting 
ordered and disordered vortex phases have 
been imaged,
and are associated with history effects and anomalous
voltage responses of the vortex lattice.

As illustrated by the lower curve 
in Fig.~\ref{fig:twait}, the waiting time $t_{w}$ decreases 
rapidly as the applied driving
force $f_{d}$ increases, until for $f_{d}>0.0065f_{0}^{*}$ 
there is no measurable
waiting time.  
Since a channel is opening along the length of the system during
the waiting time, $t_{w}$ should be proportional to the amount
of time required for vortices to move along this channel.
This is given by:
\begin{equation}
t_{w} \propto L_{x}/f_{d} ,
\end{equation}
where $L_{x}$ is the system size along the length of the channel.
A fit of $t_{w}$ to $1/f_{d}$ is shown in Fig.~\ref{fig:twait} by
the dashed line.

The time $t_{s}$ required for the response to saturate
decreases with $f_{d}$ for $f_{d}<0.0065f_{0}^{*}$, 
as indicated by the upper curve in Fig.~\ref{fig:twait}.  
For $t_{w}<t<t_{s}$, the ordered channel begins to spread
through the sample in the direction transverse to the applied driving
force.  The motion of this ordered front resembles the transverse
depinning of a longitudinally driven interface.  The faster the
interface is moving in the longitudinal direction, the lower the
transverse depinning force is, since the interface becomes
less rough.  In a model for the transverse depinning of a longitudinally
driven elastic string
\cite{transversestring}, the transverse depinning force 
$f_{dp}^{T}$ was found
to decrease as $f_{dp}^{T} \propto (f_{d} - f_{dp})^{-\alpha}$
with $\alpha = 2/3$.
In the case of the moving front of ordered vortices,
an effective transverse force $f_{\rm eff}^{T}$ on the front is provided by the
interactions between the pinned and 

\begin{figure}
\center{
\epsfxsize = 3.5in
\epsfbox{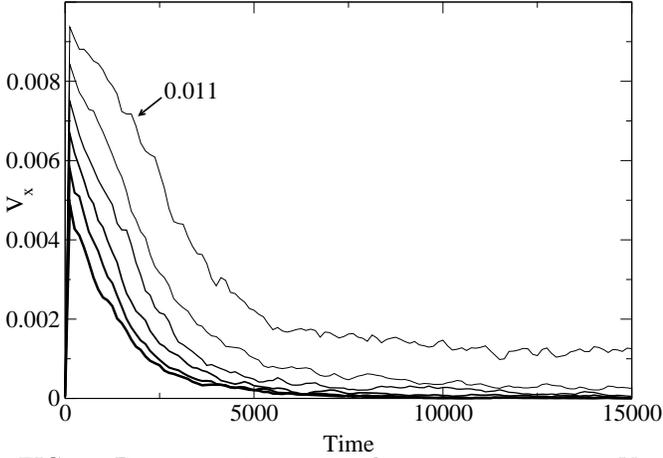}}
\caption{Decreasing
transient voltage response curves $V_{x}$ versus time for superheated
state, for $s_{m}=0.7$ and $f_{d}/f_{0}^{*}=$ (bottom) 0.006, 0.007,
0.008, 0.009, 0.010, and 0.011 (top).}
\label{fig:newfig6}
\end{figure}

\noindent
moving vortices.
If we assume that these interactions do not change with $f_{d}$, since
$f_{d}$ is not being applied in the transverse direction, then 
$f_{\rm eff}^{T}$ remains constant while $f_{dp}^{T}$ decreases with
increasing applied drive $f_{d}$.  As a result, the ordered front
propagates outward more
quickly as $f_{d}$ increases, and the saturation time $t_{s}$ required for the
ordered channel to fill the entire sample decreases.  
In Fig.~\ref{fig:twait} the dotted line indicates a fit to 
\begin{equation}
t_{s} \propto (f_{d}-f_{d}^{0})^{-1} ,
\end{equation}
with $f_{d}^{0}=0.0053$. 
For $f_{d}<0.0053f_{0}^{*}$, there was no voltage response to the applied
current over the time period we considered (indicating that $t_w>400000$)
and the vortices remained stationary in the supercooled state.
The apparent discontinuity in $t_w$ from a finite to an unmeasured value
arises only because our simulations were performed for a finite amount
of time.

\begin{figure}
\center{
\epsfxsize = 3.5in
\epsfbox{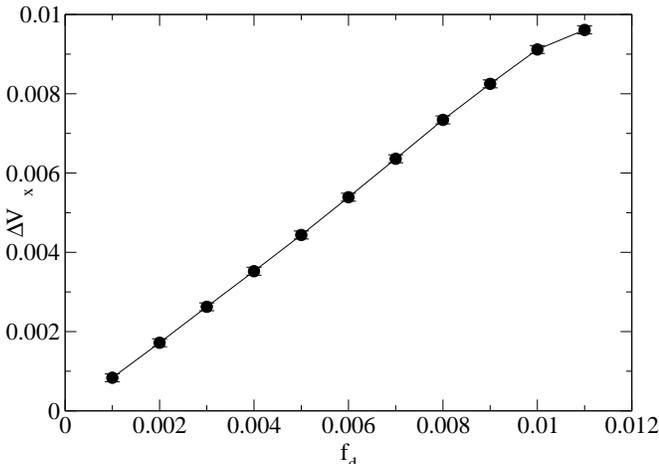}}
\caption{
Total voltage drop $\Delta V_{x}$ between the peak and plateau values
of the decreasing response curves as a function of applied drive $f_{d}$.
}
\label{fig:yield}
\end{figure}

\begin{figure}
\center{
\epsfxsize = 3.5in
\epsfbox{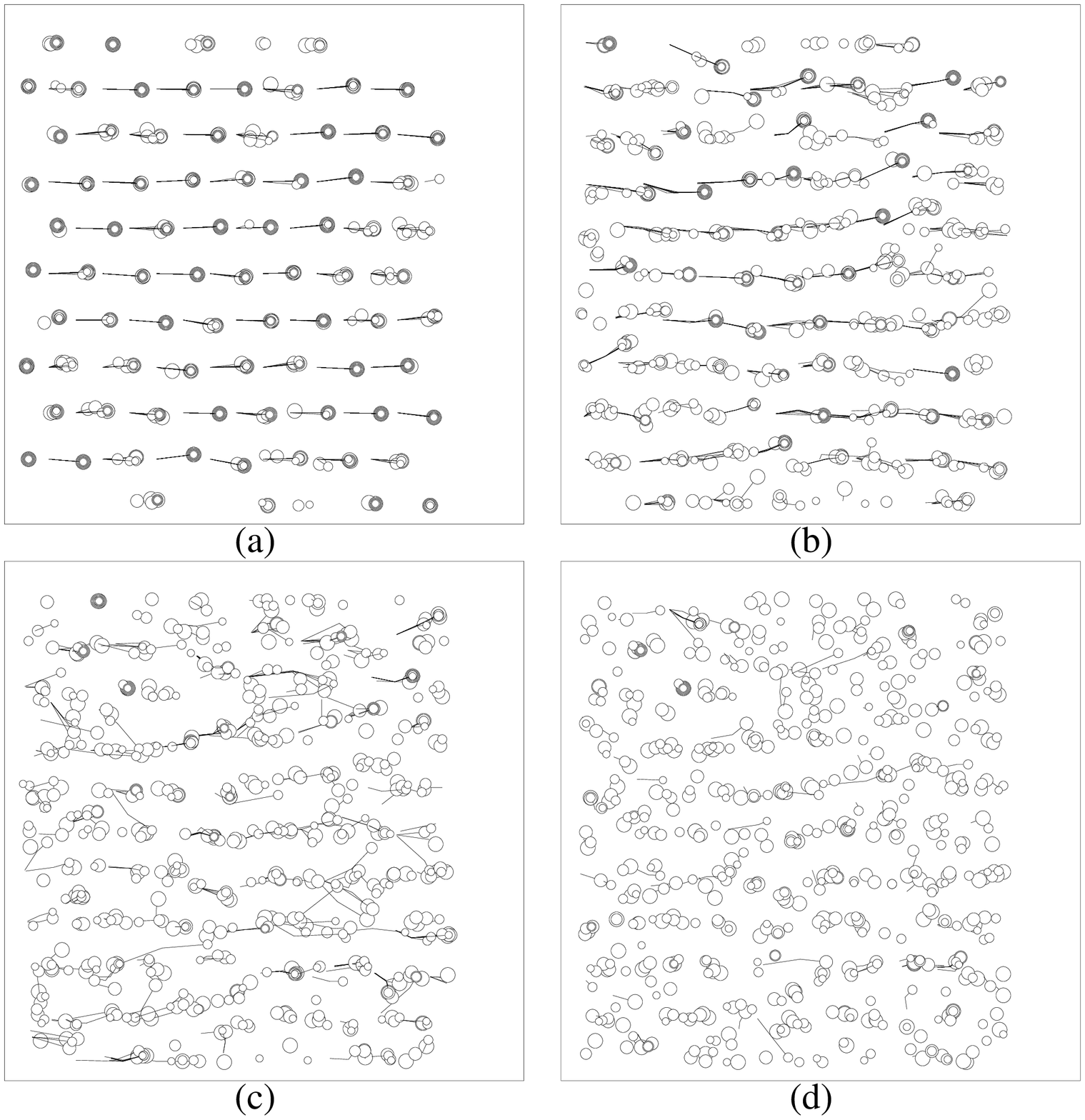}}
\caption{Vortex positions (circles)  and trajectories (lines) for the 
superheated phase at $s_m=0.7$ and $f_d=0.006 f_{0}^{*}$, at times:
(a) $t=500$; (b) $t=1500$; (c) $t=2500$; (d) $t=4500$.  Each vortex line
breaks apart into pancakes as it encounters the pinning sites.
}
\label{fig:heatI}
\end{figure}

We next consider transient effects produced by 
superheating the lattice.
In Fig.~\ref{fig:newfig6} we show a superheated system 
at $s_{m}=0.7$ prepared in the ordered state by
artificially placing the vortices 
into perfectly aligned columns 
at $t=0$. Here we find 
a large initial $V_x$
response that 
decays. 
With larger $f_{d}$ the decay {\it takes an increasingly
long time}. 
The time scale for the decay is much {\it shorter} 
than the time scale
for the increasing response in Fig.~\ref{fig:newfig2}.
As illustrated in Fig.~\ref{fig:newfig3}(c) and (d), the $z$-axis
correlation $C_{z}$ 
decays more rapidly than the overall voltage
response $V_{x}$, indicating that vortex motion continues to
occur even after the vortex lines have been broken apart into
individual pancakes.  The form of $V_{x}$ cannot be fit by any
simple function over a significant period of time.  The decay
of $C_{z}$, however, is nearly linear with time before saturating
at $t \approx 3500$.
For $f_{d} \ge 0.011f_{0}^{*}$ the voltage does not decay
completely away to zero, but the vortices continue to move in
the disordered state.

The decaying response of the superheated vortex state resembles
the yield drop curves observed in crystalline solids, as first
noted in low-T$_c$ materials by Good and Kramer \cite{Good14b}.
In a yield drop, the stress required to maintain a constant shear
strain rate decreases with time.  Such a drop occurs when the
crystalline solid has a low initial concentration of defects,
and the drop is associated with a proliferation of dislocations
inside the crystal.  In the case of the 
superheated vortex lattice, a transition from the ordered to the 
disordered state occurs and the 

\begin{figure}
\center{
\epsfxsize=3.5in
\epsfbox{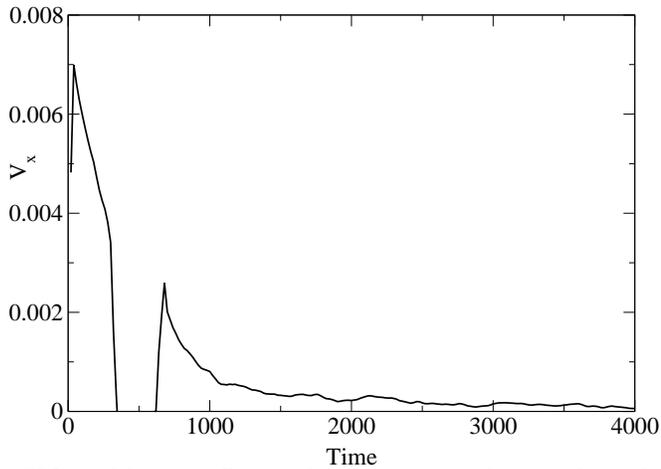}}
\caption{Memory effect in the $V_{x}$ response of a superheated
system with $s_m=0.7$ and $f_d=0.009f_{0}^{*}$.  The current is
shut off for 400 time steps beginning at $t=400$, and is
turned back on at $t=800$.}
\label{fig:memory}
\end{figure}

\noindent
vortex velocity drops.  Good and Kramer 
find that the total drop in the response of the vortex lattice increases
linearly with the applied voltage \cite{Good14b}.  To paraphrase their
argument, the voltage response of the lattice can 
be written
$V_{x}=F_{m}R_{f}f_{d}$, where $F_{m}$ is the fraction
of the vortex lattice that is moving, and $R_{f}$ is the flux
flow resistance, or the slope of the $V(I)$ curve at high currents.
The change in voltage response from the initial peak, where
$F_{m}^{0}=1$ and the entire lattice is moving, to the saturated
plateau where $F_{m}^{f}=0$ and all of the vortices have
repinned, can then be written 
\begin{equation}
V_{x}^{0}-V_{x}^{f} = (F_{m}^{0}-F_{m}^{f})R_{f}f_{d},
\end{equation}
or $\Delta V_{x} = R_{f}f_{d}$.  In Fig.~\ref{fig:yield}
we plot the difference $\Delta V_{x}$ between $V_{x}$ at the
peak and plateau as a function of applied drive $f_{d}$, and
find a linear relation with slope $R_{f}=0.93$, in good agreement
with the expected value $R_{f}=1$.  There is a deviation from
linearity for $f_{d}>0.010$ because at higher drives, 
$F_{m}^{f} > 0$ since all of the vortices do not repin.

The vortex positions and trajectories for a superheated sample with
$s_m = 0.7$ and $f_d = 0.006f_{0}^{*}$, 
as in Fig.~\ref{fig:newfig6},
are shown in Fig.~\ref{fig:heatI}(a-d).
In Fig.~\ref{fig:heatI}(a) the initial vortex state is ordered. 
In Fig.~\ref{fig:heatI}(b-d) the vortex lattice becomes
disordered and pinned in a {\it homogeneous} manner
rather 
than through nucleation.
Each vortex line is decoupled by the point pinning
as it moves 
until the entire line dissociates and is pinned. 
Since all of the vortex lines decouple simultaneously, this decaying
response process occurs much more rapidly than the increasing response
of the supercooled sample, which required nucleation and growth of
a channel.

In Fig.~\ref{fig:memory}
we demonstrate the presence of a 
{\it memory} effect 
by abruptly shutting off $f_d$ at $t=400$ for 400 time steps. The vortex 
motion stops when the drive is shut off,
and when $f_d$ is re-applied $V_x$ resumes at the 
same 

\begin{figure}
\center{
\epsfxsize = 3.5in
\epsfbox{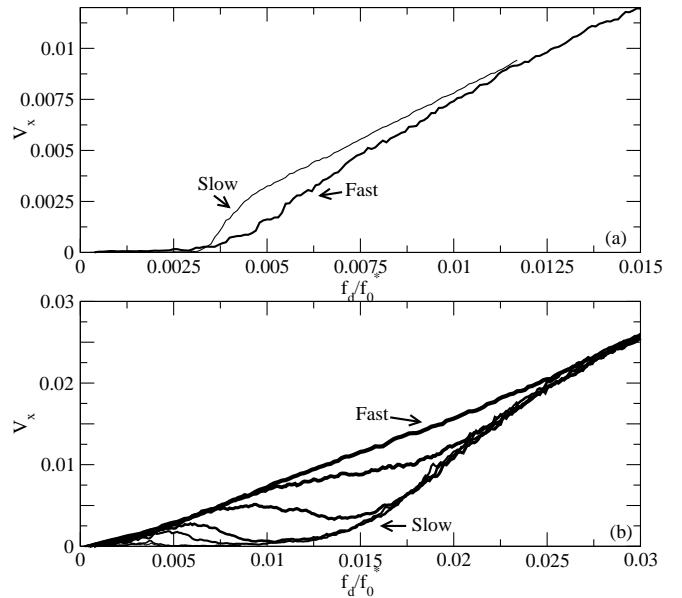}}
\caption{(a): V(I) for a supercooled system
at $s_m = 2.0$.  Top (light line): slow $\delta f_{d}$ of $0.0001 f_{0}^{*}$ 
every 2000 time steps. 
Bottom (heavy line): $\delta f_{d} = 0.001 f_{0}^{*}$. 
(b): V(I) for a superheated system at
$s_m = 0.7$.  From left to right, $\delta f_{d} = $
(fast) $0.02 f_{0}^{*}$, $0.01 f_{0}^{*}$,
$0.005 f_{0}^{*}$, $0.002 f_{0}^{*}$, $0.001 f_{0}^{*}$, 
$0.0002 f_{0}^{*}$, and 
$0.0001 f_{0}^{*}$ (slow).
}
\label{fig:ramp}
\end{figure}

\noindent
point. We find such memory on both the increasing and 
decreasing response curves.  Since we are at zero temperature, it is
the applied current, and not thermal effects,
that is responsible for the vortex motion, and thus the vortices cannot
adjust their positions in the absence of a driving current.
The response curves 
and memory effect 
seen here are very similar to those 
observed in experiments \cite{Xiao7,Good14b}.

We next consider the effect of changing the rate $\delta f_{d}$
at which the driving force 
is increased on $V(I)$ in both
superheated and supercooled systems.
Fig.~\ref{fig:ramp}(a) shows 

\begin{figure}
\center{
\epsfxsize = 3.5in
\epsfbox{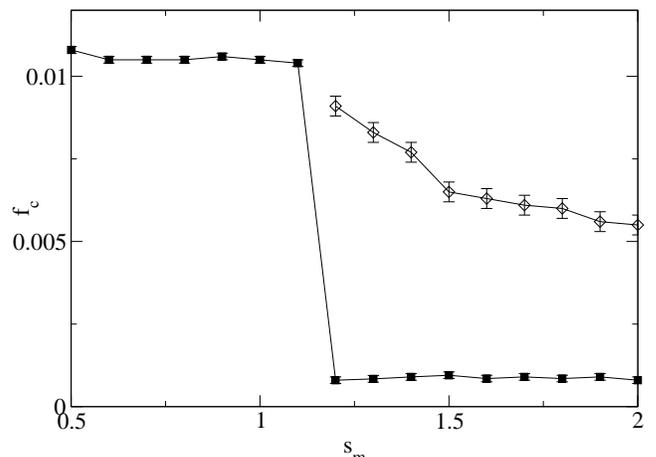}}
\caption{
Effect of supercooling on $f_{c}$.  Filled squares: equilibrium $f_{c}$.
Open diamonds: $f_{c}$ for samples prepared in a disordered, decoupled
state.}
\label{fig:disorder}
\end{figure}

\begin{figure}
\center{
\epsfxsize = 3.5 in
\epsfbox{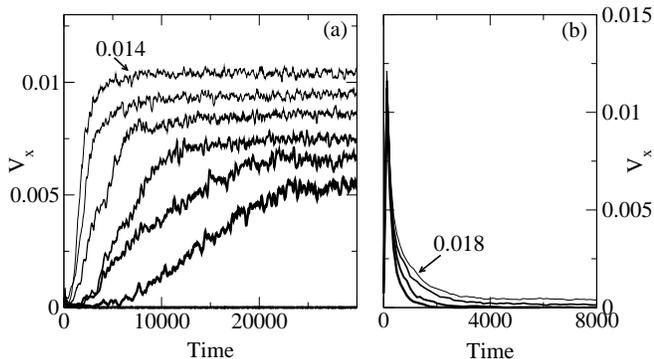}}
\caption{Transient voltage response curves $V_x$ versus time.
(a) Increasing response for
supercooled state, for $n_v= 2.5$ and $f_d/f_{0}^{*} =$ 
(bottom) 0.008, 0.009, 0.010, 0.011,
0.012, 0.013, and 0.014 (top).
(c) Decreasing response for superheated state, for $n_v=3.1$ and 
$f_d/f_{0}^{*} =$ 
(bottom) 0.0016, 0.0017, 0.00175 and (top) 0.0018.
}
\label{fig:trans}
\end{figure}

\noindent
$V_x$ versus $f_d$, which is 
analogous to a V(I) curve, for 
the 
supercooled system at $s_m=2.0$ 
prepared in a disordered 
state.  
$V_x$ remains low during a fast ramp,
when the vortices in the strongly pinned disordered state 
cannot reorganize into the more ordered state. 
There is also 
considerable hysteresis 
since the vortices reorder at higher 
drives producing a
higher value of $V_x$ during the ramp-down.
For the slower ramp the vortices have time 
to reorganize into the weakly pinned ordered state, 
and remain ordered, producing {\it no hysteresis} in 
V(I).

In a superheated sample, the reverse behavior occurs.
Fig.~\ref{fig:ramp}(b) shows V(I) curves at different $\delta f_{d}$
for a system with $s_m = 0.7$ prepared in the ordered
state.  Here,  the fast ramp has a {\it higher} value of $V_x$ 
corresponding to the ordered state
while the slow ramp has a low value of $V_x$.  
During a slow initial ramp in the superheated state the 
vortices gradually disorder through rearrangements but 
there is no net vortex flow through the sample.  Such 
a phase was proposed by Xiao {\it et al.} \cite{Xiao7}
and seen in recent experiments on BSCCO samples 
\cite{Konczykowski16}.  At the slower $\delta f_{d}$,
we find {\it negative} $dV/dI$ characteristics 
which resemble those seen in
low- \cite{Frindt14,Borka24,Alekseevskii25,Tomy25a,Tomy25b} 
and high- \cite{deGroot25} 
temperature superconductors.
Here, $V(I)$ initially increases 
as the 
vortices flow in the ordered state,
but the vortices decouple as the lattice moves,
increasing $f_c$ and dropping
$V(I)$ back to zero, 
resulting in an N-shaped 
characteristic.

To demonstrate the effect of vortex lattice
disorder on the critical current,
in Fig.~\ref{fig:disorder} we plot the equilibrium $f_c$ 
along with $f_c$ obtained
for the supercooled system, in which each sample is prepared in a state
with $s_{m} = 0.5$, and then $s_m$ is raised to a new value above 
$s_{m}^{c}$ before $f_c$ is measured.  The disorder
in the supercooled state produces a value of $f_c$ 
between the two extrema observed in the equilibrium state.
Note that the sharp transition in $f_c$
associated with equilibrium systems is now smooth.
A similar increase in the critical current when the vortex lattice
has been prepared in a disordered state, rather 

\begin{figure}
\center{
\epsfxsize = 3.5 in
\epsfbox{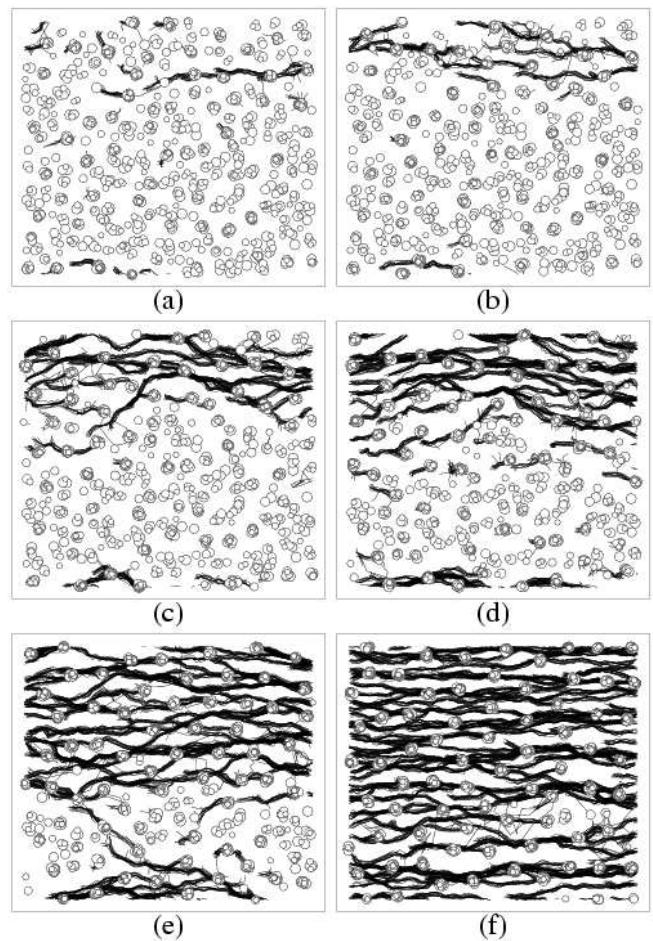}}
\caption{Pancake vortex positions (circles)  and trajectories (lines) for the 
supercooled phase
at $n_v=2.5\lambda^2$ 
and $f_d=0.009 f_{0}^{*}$, seen from the top of the sample, at
times: (a) $t=5000$; (b) $t=7500$; (c) $t=10000$; (d) $t=12500$;
(e) $t=15000$; (f) $20000$.  
An ordered channel 
forms in the sample and grows outward.}
\label{fig:cool}
\end{figure}

\begin{figure}
\center{
\epsfxsize = 3.5in
\epsfbox{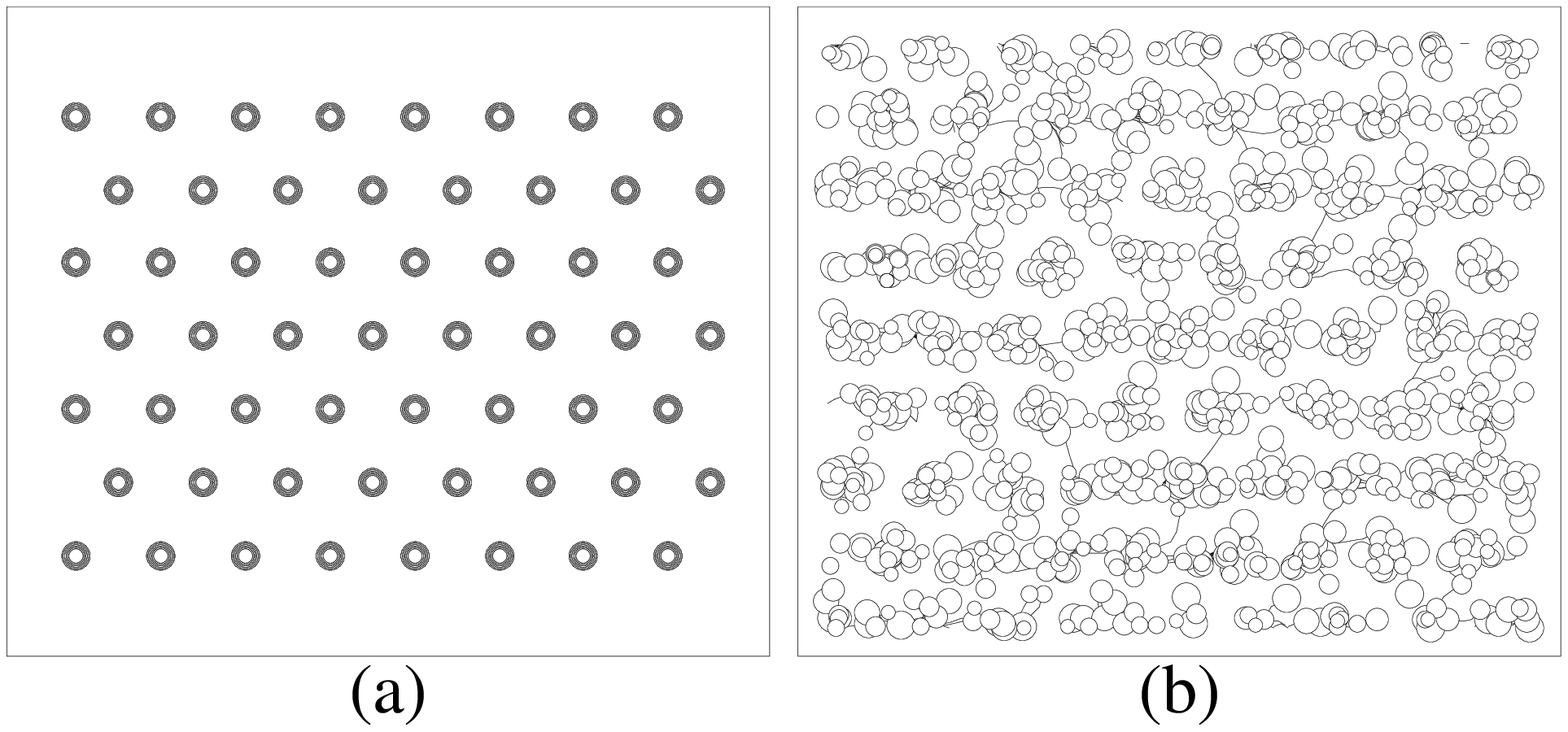}}
\caption{Vortex positions (circles)  and trajectories (lines) for the 
superheated phase at $n_v=3.1$ and $f_d=0.016 f_{0}^{*}$, at times:
(a) $t=0$; (b) $t=1500$. Each vortex line
breaks apart into pancakes as it encounters the pinning sites.
}
\label{fig:heat}
\end{figure}

\noindent
than in an ordered
state, has also been observed experimentally as a history effect
\cite{Baker14a,Steingart26,Kupfer27,Koch28,Wordenweber29,Obara30,Dilley31}.

We find the same metastable and history-dependent effects 
described above on either
side of a magnetic-field 

\begin{figure}
\center{
\epsfxsize = 3.5in
\epsfbox{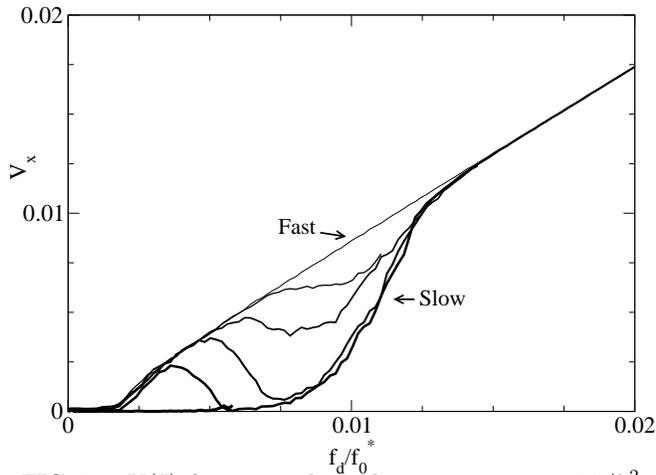}}
\caption{$V(I)$ for a superheated system at $n_{v}=3.1/\lambda^{2}$.  From left
to right, $\delta f_{d}=$ (fast) $0.01 f_{0}^{*}$, $0.005 f_{0}^{*}$,
$0.002 f_{0}^{*}$, $0.001f_{0}^{*}$, $0.0002f_{0}^{*}$, and 
$0.0001f_{0}^{*}$ (slow).}
\label{fig:vi}
\end{figure}

\noindent
driven disordering transition.  To
demonstrate this, in Fig.~\ref{fig:trans}(a) we show $V_{x}$
for several different drives $f_{d}$ for 
a sample with
dense, strong pinning ($f_{p}=0.08f_{0}^{*}$, $n_{p}=8.0/\lambda^{2}$)
at $n_{v}=2.5/\lambda^{2}$, where a field-induced decoupling transition occurs
above $n_{v}=3.0/\lambda^{2}$. 
The sample shown in Fig.~\ref{fig:trans}(a)
would be coupled in equilibrium but has
been prepared in a decoupled state.  For $f_{d} \ge 0.009f_{0}^{*}$
we observe a time dependent increasing response similar to that
shown in Fig.~\ref{fig:newfig2}.  The rising response is associated
with the formation of a channel of ordered, coupled vortices as
illustrated in Fig.~\ref{fig:cool}.  In Fig.~\ref{fig:trans}(b) we
show $V_{x}$ for the same sample in a superheated state at
$n_{v}=3.1/\lambda^{2}$, where the sample would be disordered in
equilibrium but has been prepared in an ordered state.  Here we
find a transient decreasing response as the lattice disorders,
similar to the response in Fig.~\ref{fig:newfig6}.  Images of
the sample before and after the current pulse is applied are
illustrated in Fig.~\ref{fig:heat}, showing the transition from
an ordered to a disordered state.  The presence of transients
also leads to a ramp-rate dependence of the $V(I)$ curve, and
in Fig.~\ref{fig:vi} we illustrate the appearance of N-shaped
characteristics in the superheated system when the current
is ramped at increasing rates for the same sample.

Our simulation does not contain a surface 
barrier which can inject disorder at the edges. Such an effect
is proposed to explain 
experiments in which AC current pulses induce an increasing
response as the vortices reorder but DC pulses
produce a decaying response \cite{Andrei6,Paltiel8}. 
We observe no difference between AC and DC drives. 

In low temperature superconductors, 
a rapid increase in $z$-direction vortex wandering 
occurs simultaneously with vortex disordering \cite{Ling15}, 
suggesting that the change in $z$-axis correlations may be 
crucial in these systems as well.  Our results, along with recent 
experiments
on 
layered superconductors, suggest that the transient 
response seen in low temperature materials should also appear
in layered materials.

In summary we have investigated transient and metastable states near
the 3D-2D transition by supercooling or superheating the system. We find
voltage-response curves and memory effects that are very similar to those
observed in experiments, 
and we identify the microscopic vortex dynamics associated with
these transient features.  
In the supercooled case the vortex motion 
occurs through nucleation of a channel of ordered moving vortices followed
by an increase in the channel width over time.
In the superheated case the ordered phase homogeneously disorders. We also
demonstrate that the measured critical current depends on 
the vortex lattice preparation
and on the current ramp rate.  

We acknowledge helpful discussions with 
E. Andrei, S. Bhattacharya, X.S. Ling, Z.L. Xiao, and E. Zeldov. This
work was supported by CLC and CULAR (LANL/UC) by NSF-DMR-9985978, by
the Director, Office of Adv.\ Scientific Comp.\ Res., Div.\ of Math., 
Information and Comp.\ Sciences, U.S.\ DoE contract DE-AC03-76SF00098,
and by the U.S. Dept. of Energy under Contract No. W-7405-ENG-36.

\end{document}